\documentclass[twocolumn,prb,aps,showpacs,superscriptaddress,amsfonts,amssymb,amsmath]{revtex4}
\usepackage{amsmath}
\usepackage{graphicx}
\begin{document}
\title{A direct calculation of critical exponents of two-dimensional anisotropic Ising model}
\author{Gang Xiong}
\affiliation{Physics Department, Beijing Normal University,
Beijing 100875, P. R. China}
\author{X. R. Wang}
\affiliation{Physics Department, The Hong Kong University of
Science and Technology, Clear Water Bay, Hong Kong SAR, China}

\date{Draft on \today}

\begin{abstract}
Using an exact solution of the one-dimensional (1D) quantum
transverse-field Ising model (TFIM), we calculate the critical
exponents of the two-dimensional (2D) anisotropic classical Ising
model (IM). We verify that the exponents are the same as those of
isotropic classical IM. Our approach provides an alternative means
of obtaining and verifying these well-known results.
\end{abstract}
 \maketitle

It is well known that the two-dimensional (2D) classical Ising
model (IM)\cite{ising} has a thermodynamic continuous phase
transition (CPT) from (anti-)ferromagnetic phase to paramagnetic
phase at a critical temperature. Later it was also known\cite
{fradkin,suzuki} that a one-dimensional (1D) quantum
transverse-field Ising model (TFIM) can be mapped into the 2D
anisotropic classical IM under certain limit. Thus, a connection
between the ground state of 1D quantum TFIM and thermodynamical
equilibrium state of a 2D classical IM was established. So far,
our knowledge about the thermodynamical properties of 2D IM,
especially those about critical phenomena, were mainly obtained by
some sophisticate techniques such as high-temperature and
low-temperature expansion\cite{domb}, renormalization group
calcualtion\cite{ma}, Onsager's exact matrix-calculation and its
variations\cite{onsager,yang}. Normally, these techniques are
computational involving, and it is often not easy to be understood
by the beginners and people outside of the field. In this short
paper, we would like to introduce a direct way to obtain the
quantities like critical exponents of 2D IM. The method is based
on the exact solution of the ground state of 1D
TFIM\cite{barouch}, which, in turn, can be obtained by using a
simple Jordan-Winger transformation\cite{jordan}.

The Hamiltonian of a 2D classical IM is\cite{ising}
\begin{eqnarray}
    H=-J_x\sum_{x,y}{\sigma_{x,y}\sigma_{x+1,y}}-J_y\sum_{x,y}{\sigma_{x,y}\sigma_{x,y+1}}
    \label{2DCIM}
\end{eqnarray}
where $\sigma_{x,y}$ is the classical Ising spin on site $(x,y)$
with two possible values $\pm1$, and $J_x$ and $J_y$ are spin-spin
couplings along $x-$axis(row) and $y-$axis(column) directions,
respectively. It is well known that this model has a thermodynamic
CPT at a critical temperature $T=T_c$\cite{onsager,yang}. Later it
was shown\cite{fradkin,suzuki} that with the relations and limits
\begin{eqnarray}
    \frac{J_y}{k_BT}\rightarrow +\infty\nonumber\\
    \frac{J_x}{k_BT}\equiv\frac{J}{h}e^{\frac{-2J_y}{k_BT}}\rightarrow0\nonumber\\
    \tau \equiv\frac{1}{h}e^{\frac{-2J_y}{k_BT}}\rightarrow0
    \label{limits}
\end{eqnarray}
the transfer matrix between two neighboring columns
\begin{eqnarray}
    \hat{T}\approx\hat{1}-\tau\hat{H}
    \label{map}
\end{eqnarray}
where $\hat{1}$ is identity operator and
\begin{eqnarray}
    \hat{H}=-J\sum_n{\hat{\sigma}_n^x\hat{\sigma}_{n+1}^x}-h\sum_n{\hat{\sigma}_n^z}
    \label{1DTFIM}
\end{eqnarray}
is the Hamiltonian operator of a 1D quantum TFIM where
$\hat{\sigma}_n^x$ and $\hat{\sigma}_n^z$ are standard Pauli
operators on site $n$. Thus the 2D anisotropic classical IM is
mapped into a 1D quantum TFIM with the following correspondences
between the two models\cite{fradkin} : (1) the equilibrium state
of 2D IM vs. the ground state of 1D TFIM; (2) the free energy of
2D IM vs. the ground-state energy of 1D TFIM; (3) the ensemble
averages of physical quantities of 2D IM vs. the ground-state
expectation values of time-ordered operators of 1D TFIM; (4) the
temperature $T$ of 2D IM vs. the transverse field $h$ of 1D TFIM.

With the so-called hard-core boson transformation
\begin{eqnarray}
    \hat{\sigma}_n^x=\hat{b}_n^{\dag}+\hat{b}_n \nonumber\\
    \hat{\sigma}_n^y=i(\hat{b}_n^{\dag}-\hat{b}_n) \nonumber\\
    \hat{\sigma}_n^z=2\hat{b}_n^{\dag}\hat{b}_n-1,
    \label{boson}
\end{eqnarray}
the Jordan-Wigner transformation\cite{jordan}
\begin{eqnarray}
    \hat{b}_n=\hat{c}_ne^{-i\pi\sum_{l=1}^{n-1}{\hat{c}_l^{\dag}\hat{c}_l}}
    \nonumber\\
    \hat{b}_n=\hat{c}^{\dag}_ne^{i\pi\sum_{l=1}^{n-1}{\hat{c}_l^{\dag}\hat{c}_l}}
    \label{jw}
\end{eqnarray}
and the Fourier transformation
\begin{eqnarray}
    \hat{c}_n=\frac{1}{\sqrt{2\pi}}\int_{-\pi}^{\pi}dk\hat{c}_ke^{-ikn}\nonumber\\
    \hat{c}^{\dag}_n=\frac{1}{\sqrt{2\pi}}\int_{-\pi}^{\pi}dk\hat{c}^{\dag}_ke^{ikn}\nonumber\\
    \label{fourier}
\end{eqnarray}
the Hamiltonian of 1D TFIM is rewritten into
\begin{eqnarray}
    \hat{H}=Nh+\int_{0}^{\pi}dk\hat{H}_k\nonumber\\
    \hat{H}_k=2iJ\sin{k}(\hat{c}^{\dag}_k\hat{c}^{\dag}_{-k}+\hat{c}_{-k}\hat{c}_k)\nonumber\\
    -2(h+J\cos{k})(\hat{c}^{\dag}_k\hat{c}_k+\hat{c}^{\dag}_{-k}\hat{c}_{-k})
\end{eqnarray}
where $N$ is site number,
$\hat{c}_n$,$\hat{c}^{\dag}_n$,$\hat{c}_k$,$\hat{c}^{\dag}_k$, are
fermion operators. Denote $\left|0\right>$ as fermions' vacuum
state, the ground state and ground-state energy of $\hat{H}$ are
obtained as\cite{barouch}
\begin{eqnarray}
    \left|\Phi_G\right>=\prod_{k=0}^{\pi}(u_k\hat{c}^{\dag}_k\hat{c}^{\dag}_{-k}+v_k)\left|0\right>\nonumber\\
    E_G=Nh+\int_{0}^{\pi}dke_G(k)
    \label{ground1}
\end{eqnarray}
where
\begin{eqnarray}
    u_k=\frac{-e_G(k)}{\left[e_G^2(k)+4J^2\sin^2k\right]^{1/2}}\nonumber\\
    v_k=\frac{2iJ\sin{k}}{\left[e_G^2(k)+4J^2\sin^2k\right]^{1/2}}\nonumber\\
    e_G(k)=-2\left(h+J\cos{k}+\sqrt{h^2+2Jh\cos k+J^2}\right).
    \label{ground2}
\end{eqnarray}
Denote
\begin{eqnarray}
    t=\frac{h}{|J|},
\end{eqnarray}
the excitation gap of $\hat{H}$ is
\begin{eqnarray}
    \Delta_E=2|J||t-1|.
\end{eqnarray}
Thus the critical point of the ground-state CPT of the 1D TFIM is
at $t=t_c=1$ where $\Delta_E$ vanishes.

Now that the ground state of 1D quantum TFIM is obtained, critical
behaviors of physical quantities of the ground state can be
calculated directly. Let us first consider the `heat capacitance'.
According to the correspondences between 2D IM and 1D TFIM, the
`heat capacitance' in 1D TFIM is
\begin{eqnarray}
    C_{TFIM}=-h\frac{\partial^2E_G}{\partial h^2}.
\end{eqnarray}
Put in the ground-state energy obtained in Eq.(\ref{ground1}) and
(\ref{ground2}), we have
\begin{eqnarray}
    C_{TFIM}=t^{-1/2}\int_{0}^{\pi}\frac{\sin^2kdk}
    {[g^2(t)+4\cos^2\frac{k}{2}]^{3/2}}
\end{eqnarray}
where
\begin{eqnarray}
    g(t)\equiv(t-t_c)/\sqrt{t}
\end{eqnarray}
Near the critical point, i.e., $t\approx t_c=1$, the denominator
in the integral can be expanded as
\begin{eqnarray}
    \left[g^2(t)+4\cos^2\frac{k}{2}\right]^{3/2}\approx8\cos^3\frac{k}{2}+3g^2(t)\cos\frac{k}{2}.
\end{eqnarray}
After some calculations one can obtain
\begin{eqnarray}
    C_{TFIM}\approx\left(\frac{1}{2}\ln\frac{32}{3}-1\right)-\ln|t-t_c|.
\end{eqnarray}
Therefore, the critical behavior of the heat capacitance is the
same as that of 2D isotropic classical IM\cite{onsager}.

The magnetization along the spin-spin interaction direction in 1D
TFIM is defined as
\begin{eqnarray}
    M_x=\frac{1}{N}\sum_{n=1}^{N}{\left<\Phi_G|\hat{\sigma}_n^x|\Phi_G\right>}
    =\left<\Phi_G|\hat{\sigma}_n^x|\Phi_G\right>.
    \label{mag}
\end{eqnarray}
However, there are two facts which make it unable to obtain the
correct value of $M_x$ directly. One is that the 1D TFIM in
Eq.(\ref{1DTFIM}) has no magnetic field component along the
spin-spin interaction direction so that $M_x$ defined in
Eq.(\ref{mag}) is always zero. The other is that
$\hat{\sigma}_n^x$ in fermion representation is a non-local
operator so that it is very difficult to calculate its expectation
value even when the ground state of 1D TFIM with non-zero magnetic
field component along the spin-spin interaction direction is
obtained. Fortunately, the magnetization can be extracted from the
ground-state spin-spin correlation function defined
as\cite{barouch}
\begin{eqnarray}
    G_{xx}(r)=\left<\Phi_G|\hat{\sigma}_n^x\hat{\sigma}_{n+r}^x|\Phi_G\right>.
    \label{total}
\end{eqnarray}
This correlation contains two parts of contributions. The first
part is a long-range correlation of constant value due to the
long-range magnetic order of the ground state, and the amplitude
of this part is the square of the magnetization. The second part
is the correlation of quantum fluctuations, and the amplitude of
this part goes to zero in the limit $r\rightarrow\infty$. Thus the
magnetization equals to the square root of the amplitude of the
correlation function in Eq.(\ref{total}) in the limit
$r\rightarrow\infty$. The spin-spin correlation function of 1D
TFIM was evaluated extensively in Barouch and McCoy's
paper\cite{barouch}, where the correlation function was expressed
in the form of a T$\ddot{o}$plitz determinant and its asymptotic
behaviors at large yet finite distance $r$ were calculated with
the use of a theorem of Szego. One exact result is\cite{barouch}
\begin{eqnarray}
    G_{xx}(r\rightarrow\infty)=[sign(J)]^r(1-t^2)^{1/4}
    \label{correl}
\end{eqnarray}
where $sign(J)$ is the sign of $J$. Here we provide a direct way
to obtain this result. Let us consider a finite chain of $N$ spins
with free boundary condition and calculate the correlation
function of the first spin and the $N-$th spin, i.e., the
ground-state expectation value of
$\hat{\sigma}_1^x\hat{\sigma}_N^x$ in the limit
$N\rightarrow\infty$. With Eq.(\ref{boson}), Eq.(\ref{jw}) and
Eq.(\ref{fourier}), it is easy to show
\begin{eqnarray}
    <\Phi_G|\hat{\sigma}_1^x\hat{\sigma}_N^x|\Phi_G>
    =<\Phi_G|(\hat{b}_{1}^{\dag}+\hat{b}_{1})
    (\hat{b}_{N}^{\dag}+\hat{b}_{N})|\Phi_G>\nonumber\\
    =<\Phi_G|(\hat{c}_{1}^{\dag}+\hat{c}_{1})
    (\hat{c}_{N}^{\dag}-\hat{c}_{N})e^{i\pi\sum_{l=1}^{N}
    \hat{c}_{l}^{\dag}\hat{c}_{l}}|\Phi_G>\nonumber\\
    =<\Phi_G|(\hat{c}_{1}^{\dag}+\hat{c}_{1})(\hat{c}_{N}^{\dag}-\hat{c}_{N})
    e^{i\pi\int_{0}^{\pi}dk
    (\hat{c}^{\dag}_{k}\hat{c}^{\dag}_{k}+\hat{c}^{\dag}_{-k}\hat{c}^{\dag}_{-k})}
    |\Phi_G>\nonumber\\
    =<\Phi_G|(\hat{c}_{1}^{\dag}+\hat{c}_{1})(\hat{c}_{N}^{\dag}-\hat{c}_{N})|\Phi_G>
\end{eqnarray}
with the use of
\begin{eqnarray}
    \exp[i\pi(\hat{c}_{k}^{\dag}\hat{c}_{k}+\hat{c}_{-k}^{\dag}\hat{c}_{-k})]
    (u_k\hat{c}^{\dag}_k\hat{c}^{\dag}_{-k}+v_k)|0>\nonumber\\
    \equiv(u_k\hat{c}^{\dag}_k\hat{c}^{\dag}_{-k}+v_k)|0>.
\end{eqnarray}
Thus the rest work is to transform
$(\hat{c}_{1}^{\dag}+\hat{c}_{1})(\hat{c}_{N}^{\dag}-\hat{c}_{N})$
into $k-$space representation by Eq.(\ref{fourier}) and calculate
its ground-state expectation value. From the form of $|\Phi_G>$ in
Eq.(\ref{ground1}), it is obvious that only the following four
kinds of terms have non-zero expectation values
\begin{eqnarray}
    <\Phi_G|\hat{c}_k\hat{c}^{\dag}_k|\Phi_G>=1-u_k^2\nonumber\\
    <\Phi_G|\hat{c}^{\dag}_k\hat{c}_k|\Phi_G>=u_k^2\nonumber\\
    <\Phi_G|\hat{c}^{\dag}_k\hat{c}^{\dag}_{-k}|\Phi_G>=u_k v_k\nonumber\\
    <\Phi_G|\hat{c}_{-k}\hat{c}_{k}|\Phi_G>=u_k v^{\ast}_k
\end{eqnarray}
After doing some integrals, one finally obtains that in the
thermodynamic limit $N\rightarrow\infty$
\begin{eqnarray}
    <\Phi_G|\hat{\sigma}_1^x\hat{\sigma}_N^x|\Phi_G>=[sign(J)]^N(1-t^2)^{1/4}
\end{eqnarray}
for $t<t_c=1$ and is zero for $t>t_c$, which is the same as
Eq.(\ref{correl}). From this result, one can see that
\begin{eqnarray}
    M_x=(1-t^2)^{1/8}\approx(t_c-t)^{1/8}
\end{eqnarray}
for $t_c-t\rightarrow+0$ and is zero for $t>t_c$, which is exactly
the same as the critical behavior of the magnetization of 2D
isotropic IM\cite{yang}. Thus we have found that the critical
behaviors of the heat capacitance and the magnetization of 1D TFIM
and 2D classical isotropic IM are the same. According to the
scaling laws of critical phenomena\cite{ma}, we can conclude that
all the six critical exponents of the two models are equal. Thus
our approach serves as an alternative way to obtain and verify the
well-known results of 2D isotropic IM. In fact, another exact
result obtained by Barouch and McCoy\cite{barouch}, the value at
the transition point $t=t_c$ which decays as a power-low function
of $r$,
\begin{eqnarray}
    G_{xx}(t=t_c,r)=[sign(J)]^rr^{-1/4}e^{1/4}2^{1/12}A^{-3}\propto r^{-1/4}\nonumber\\
\end{eqnarray}
where $A=1.282427130$ is Glaisher's constant, was shown to be the
same as that of 2D isotropic classical IM. The exponential decay
of the quantum-fluctuation-induced spin-spin correlation around
the transition point was also shown to be equivalent to that in 2D
isotropic IM\cite{barouch}. It should be noted that it is unable
to obtain the critical behaviors of zero-field susceptibility
$\chi$ and the magnetization at the critical point under vanishing
field because the 1D TFIM in Eq.(\ref{1DTFIM}) has no magnetic
field component along the spin-spin interaction direction.

We would like to make some discussions about the above results.
From the mapping between 2D classical IM and 1D TFIM, one can see
that the approximation in Eq.(\ref{map}) is valid only when the
couplings along one direction, say, $J_y$, goes to {\it positive
infinity} while the couplings along the other direction, say,
$J_x$, goes to {\it zero}. This means that a 1D TFIM is mapped
into a 2D classical IM with {\it extremely anisotropic} couplings.
Furthermore, a 1D TFIM with {\it anti-ferromagnetic} coupling,
i.e., $J<0$, should correspond to a 2D classical IM with {\it
infinitely large ferromagnetic} coupling along one direction and
{\it vanishing anti-ferromagnetic} coupling along the other
direction. Thus the equivalence of the critical behaviors of 1D
TFIM and 2D classical isotropic IM shows that the critical
behaviors of 2D classical IM is very robust.

In summary, with the use of the exact solution of the 1D TFIM, we
calculate the critical exponents of the 2D anisotropic classical
IM and verify that the exponents are the same as those of 2D
isotropic classical IM. This shows that the universality class of
critical behaviors in 2D classical IM is quite robust, since the
2D classical IM corresponding to a 1D TFIM is extremely
anisotropic. Our approach provides an alternative means to obtain
and verify those well-known results.

GX acknowledges the support of CNSF under grant No. 10347101 and
the Grant from Beijing Normal University. XRW is supported by UGC
grants, Hong Kong.

\end{document}